\documentclass[aps,prl,twocolumn,groupedaddress,showpacs]{revtex4}

\def\bg{{\bf g}}
\def\AB{$A\&B$}
\def\rA{$\rho_A$}
\def\mrA{\rho_A}
\def\nl{\hfil\break}

\def\ctb{\cite{buz1}}

\begin{document}

\title{Comment on No-Signaling Condition and Quantum Dynamics}

\author{Pavel B\'ona}
\email{bona@fmph.uniba.sk}
\affiliation{Department of Theoretical Physics\\ FMPI, Comenius University\\
Bratislava, Slovakia}


\date{\today}


\pacs{03.65.Ud,03.65.Ta,05.20.Gg}


{\large\bf Comment on ``No--Signaling Condition and Quantum
Dynamics''} \vskip.3cm\ In carefully worded paper \cite{buz1}, the
authors tried to derive linearity (i.e. affinity on `density
matrices' =: DM) and complete positivity (CP) of general quantum
mechanical dynamics \bg\ from usual (nonrelativistic) kinematics
of quantum mechanics (QM), and from an additional ``no--signaling
condition'' (NS). I shall try to show here that the declared goals
of \cite{buz1} were not attained there.

The authors consider a given system $A$ in an arbitrary state
described by a DM $\rho_A$, as a subsystem of a composed system
$A\& B$ occuring in a pure state $|\psi_{AB}\rangle$.  The
subsystems $A$ and $B$ are spacelike separated. Different convex
decompositions of the reduced DM $\rho_A =\sum_kq_k\rho_k$ are
obtained by different choices of discrete measurements on $B$.
`Measurements' of $\lambda I_B$ give trivial decomposition of \rA,
other non-maximal measurements give decompositions of \rA\ to
density matrices. The pure-state decompositions (corresponding to
maximal measurements) are interpreted in \ctb\ as representing the
corresponding different ``probabilistic mixtures'' (PM) in the
sense of (classical) statistical ensembles (of quantal systems),
sometimes in literature called {\it Gemenge}, or also {\it genuine
mixtures} in \cite{bon}.

The time evolution transformation \bg\ (``not necessarily
linear'') of $A$ ``is {\it a priori}\ {\bf defined only on pure
states}\dots'' \footnote{Boldface in quotations are my emphases.
P.B.}, ${\bf g}: P_\psi\mapsto{\bf g}(P_\psi)$. An explicit
extension of \bg\ to all considered states of $A$, e.g. to all
decompositions $\{\rho_k,q_k\}$ of \rA, is essential, however, for
the forthcoming discussion: Effects of any deterministic (no
collapse!) semi-group of time transformations \bg\ are supposed to
be uniquely determined in QM by its initial conditions
\big[``$\bg(P_\psi)$ does not have to be a pure state'' of $A$
(!)\big]. The following observation will also support my
criticism:\nl (*) ``\dots the results of measurement on $A$ will
be completely determined by the reduced DM of the system.''
\cite[pp.2-3]{buz1}.

 Decisive for proving linearity of \bg\ is:
(**) ``\dots{\it every} PM of pure states corresponding to the DM
$\rho_A$ {\bf can be prepared} via appropriate measurements on
$B$'' (this is supported by calculations of probabilities at $A$
conditioned by results of measurements on $B$); such a process is
classified in
 \cite{EPR} as the ``reduction of the wave packet'', i.e. a use of
the projection postulate (having an {\it ontological meaning}),
what is, however, strongly rejected in\nl \cite[pp. 1 and
2]{buz1}. The linearity of \bg\ is then implied by:\nl
\hspace*{1.2cm}$\bg(\mrA) = \bg(\{\rho_k,q_k\})=\sum_j p_j
\bg(P_{\psi_j}),$ \qquad\quad ($\dagger$)\nl \big($\bg(\mrA)
:={\mrA}'(\{P_{\psi_j},p_j\})$ in \ctb\big), if valid for
arbitrary (or at least pure) decompositions \rA\ = $\sum_k
q_k\rho_k=\sum_j p_j P_{\psi_j}$; ($\dagger$) was deduced in \ctb\
from (**), and from a use of NS.

My criticism is concentrated to two points, i.e., mainly, to
(first) criticism of the way of the deduction of the restriction
($\dagger$) imposed on $\bg(\{\rho_k,q_k\})$, leading to linearity
of \bg, and to, less important, \nl (second) criticism of the
statement of the implication:\ \{{\it linearity}\ \&\ {\it
positivity (of each time evolution)\} $\Rightarrow$ \{complete
positivity of \bg}\}.

(first): The necessity of ($\dagger$) in \ctb\ is given by mere
``statics'' of \ctb, {\bf without NS}, since that kinematics
(embracing all `state space points $\bullet$' appearing as initial
conditions for \bg, and also its values $\bg(\bullet)$) does not
contain in \ctb\ any means (i.e. corresponding observables) to
ascertain locally a distinction between different kinds of
interpretation of \rA, cf. (*); then the value of \bg(\rA) should
be here the same for \rA\ considered as an indecomposable quantity
describing a quantum state of each single system $A$ in an
ensemble of equally prepared couples \AB, as well as for \rA\
representing a specific ensemble of subsystems $A$ each of which
being in one of the states $\rho_k$ taken from the set composing
the chosen convex decomposition $\{\rho_k,q_k\}$ of \rA.

It can be introduced, however, a state space for $A$ (as it was
partly done implicitly in \ctb){\it consisting of all probability
measures on density matrices} (interpreted as corresponding PM's,
and encompassing different decompositions of the same density
matrix as different points) with observables distinguishing them;
let us define then $\bg(\{\rho_k,q_k\}):=\sum_kq_k\bg(\rho_k)$ for
the case of PM $\{\rho_k,q_k\}$, and let \bg($\rho$) be
`independently' given for any (not decomposed!) density matrix
$\rho$, cf. \cite[2.1-e]{bon}. Then the proof of linearity of \bg\
in \ctb\ (with a use of NS) depends on possibility of an empirical
check of (**) (i.e. of existence of {\it physical} differences
between different decompositions of \rA\ {\bf at the instant of
the measurements on $B$}) without a use of results of measurements
on $B$. Its negative result (due to NS) does not imply
($\dagger$): All the physically indistinguishable ``at a distance
prepared PM's'' are described by \rA\ and all of them evolve to
$\bg(\mrA)\equiv\bg\left(\sum_kq_k\rho_k\right)\ 
\big[\not\equiv\bg\left(\{\rho_k,q_k\}\right)$ for nonlinear
\bg\big].

(second): Assuming linearity and positivity of each physical time
evolution transformation \bg, authors infer CP of \bg\ by applying
these properties to extensions \AB\ of the considered system $A$.
Their arguments consist, however, of a
 rephrasing of the definition of CP
and of its physical motivation published in \cite[Sec. 9.2]{dav}.

My conclusion is that the authors did not succeed in their effort
to prove in \ctb\ effectiveness of new quantummechanical axiom
called the ``no--signaling condition'', and the declared aims of
the paper \cite{buz1} were not achieved.




\vskip1.5cm

 \noindent Pavel B\'ona

{\small Department of Theoretical Physics FMPI,

Comenius University, 842 48 Bratislava, Slovakia.}


\end{document}